\documentclass[runningheads]{svmult}

\usepackage{makeidx}   
\usepackage{graphicx}  
\usepackage{subeqnar}  
\usepackage{multicol}  
\usepackage{physprbb}  
\makeindex             

\def\lsim{\hbox{ \rlap{\raise 0.425ex\hbox{$<$}}\lower 0.65ex\hbox{$\sim$} }}
\def\gsim{\hbox{ \rlap{\raise 0.425ex\hbox{$>$}}\lower 0.65ex\hbox{$\sim$} }}
\def\etal{{\it et al. }}

\begin{document}

\title*{The GRB Host Galaxies and Redshifts 
\footnote{Based in part on the observations obtained at the W.M.~Keck 
Observatory, operated by the California Association for Research in Astronomy,
a scientific partnership among Caltech, the Univ. of California and NASA;
and with the NASA/ESA Hubble Space Telescope, operated by the AURA, Inc., 
under a contract with NASA.} 
}
\toctitle{The GRB Host Galaxies and Redshifts}
\titlerunning{Host Galaxies and Redshifts}
\author{S.G. Djorgovski\inst{1}
\and S.R. Kulkarni\inst{1}
\and J.S. Bloom\inst{1}
\and D.A. Frail\inst{2}
\and F.A. Harrison\inst{3}
\and T.J. Galama\inst{1}
\and D. Reichart\inst{1}
\and S.M. Castro\inst{1}
\and D. Fox\inst{1}
\and R. Sari\inst{4}
\and E. Berger\inst{1}
\and P. Price\inst{3}
\and S. Yost\inst{3}
\and R. Goodrich\inst{5}
\and F. Chaffee\inst{5}
}
\authorrunning{Djorgovski et al.}
\institute{Palomar Observatory, Caltech, Pasadena, CA 91125, USA
\and NRAO/VLA, Socorro, NM 87801, USA
\and Space Radiation Laboratory, Caltech, Pasadena, CA 91125, USA
\and Theoretical Astrophysics, Caltech, Pasadena, CA 91125, USA
\and W.M. Keck Observatory, CARA, Kamuela, HI 96743, USA
}

\maketitle              

\begin{abstract}
Observations of GRB host galaxies and their environments in general can
provide valuable clues about the nature of progenitors. 
Bursts are associated with faint, $\langle R \rangle \sim 25$ mag, galaxies at
cosmological redshifts, $\langle z \rangle \sim 1$. 
The host galaxies span a range of luminosities and morphologies,
and appear to be broadly typical for the normal, 
evolving, actively star-forming galaxy
populations at comparable redshifts and magnitudes, but may have somewhat
elevated SFR per unit luminosity.  There are also spectroscopic hints of
massive star formation, from the ratios of [Ne III] and [O II] lines.
The observed, unobscured star formation rates are typically a few $M_\odot$/yr,
but a considerable fraction of the total star formation in the hosts may be 
obscured by dust.
A census of detected optical afterglows provides a powerful new handle on the
obscured fraction of star formation in the universe; the current results
suggest that at most a half of the massive star formation was hidden by dust.
\end{abstract}

\section{Introduction}

Host galaxies of GRBs serve a dual purpose:  determination of the redshifts,
which are necessary for a complete physical modeling of the bursts,
and to provide some insights about the possible nature of the progenitors,
e.g., their relation to massive star formation, etc.

Table 1 summarizes the host galaxy magnitudes and redshifts known to us as
of mid-June 2001.  The median apparent magnitude is $R = 24.8$ mag, with
tentative detections or upper limits reaching down to $R \approx 29$ mag.
Down to $R \sim 25$ mag, the observed distribution is consistent with deep
field galaxy counts \cite{shyc95}, but fainter than that, selection effects may
be playing a role.  We note also that the observations in the visible probe the
UV in the restframe, and are thus especially susceptible to extinction. 

\begin{table}
\caption{GRB Host Galaxies and Redshifts (June 2001)}
\begin{center}
\renewcommand{\arraystretch}{1.4}
\setlength\tabcolsep{5pt}
\begin{tabular}{llllll}
\hline\noalign{\smallskip}

  GRB &    
  $R$ mag &    
  Redshift &    
  Type $^a$ &   
  References $^b$ \\ 

\noalign{\smallskip}
\hline
\noalign{\smallskip}

 970228     &   25.2 &  0.695   & e   & \cite{fpt+99,bdk01} \\
 970508     &   25.7 &  0.835   & a,e & \cite{mdk+97,bdkf98,fpg+00} \\
 970828     &   24.5 &  0.9579  & e   & \cite{sgd+01b} \\
 971214     &   25.6 &  3.418   & e   & \cite{kdr+98,odk+98} \\
 980326     &   29.2 &$\sim$1?  &     & \cite{bkd+99}, GCN 1029\\
 980329     &   27.7 &$<$3.9    & (b) & GCN 481, 778\\
 980425 $^c$&   14   &  0.0085  & a,e & \cite{gvv+98} \\
 980519     &   26.2 &          &     & \cite{hpja99} \\
 980613     &   24.0 &  1.097   & e   & \cite{sgd+01a} \\
 980703     &   22.6 &  0.966   & a,e & \cite{dkb+98} \\
 981226     &   24.8 &          &     & \cite{fkb+99} \\
 990123     &   23.9 &  1.600   & a,e & \cite{kdo+99,bod+99} \\
 990308 $^d$&$>$28.5 &          &     & GCN 726\\ 
 990506     &   24.8 &  1.30    & e   & \cite{tbf+00,bfs01} \\
 990510     &   28.5 &  1.619   & a   & \cite{vfk+01}, GCN 757\\
 990705     &   22.8 &  0.86    & x   & \cite{mpp+00,pgg+00} \\
 990712     &   21.8 &  0.4331  & a,e & \cite{hhc+00a,hhc+00b,vfk+01} \\
 991208     &   24.4 &  0.7055  & e   & \cite{csg+01}, GCN 475, 481 \\
 991216     &   24.85&  1.02    & a,x & \cite{pgg+00}, GCN 496, 751\\
 000131     &$>$25.7 &  4.50    & b   & \cite{ahp+00} \\
 000214     &        &0.37--0.47& x   & \cite{apv+00} \\
 000301C    &   28.0 &  2.0335  & a   & \cite{jfg+01}, GCN 603, 605, 1063\\
 000418     &   23.9 &  1.1185  & e   & GCN 661, 733\\
 000630     &   26.7 &          &     & GCN 1069\\
 000911     &   25.0 &  1.0585  & e   & \cite{pap+01} \\
 000926     &   23.9 &  2.0369  & a   & GCN 851, 871\\
 010222     &$>$24   &  1.477   & a   & \cite{jpg+01}, GCN 965, 989, 1002\\

\\ 

\hline \\
\end{tabular}

{\small
~\\
\textsc{Notes}: \\
$^a$ e = line emission, a = absorption, b = continuum break, x = x-ray \\
$^b$ GCN circulars available at http://gcn.gsfc.nasa.gov/gcn/gcn3\_archive.html \\
$^c$ Association of this galaxy/SN/GRB is somewhat controversial \\
$^d$ Association of the OT with this GRB may be uncertain \\
}

\end{center}
\label{Tab1a}
\end{table}

\begin{figure}[b]
\begin{center}
\includegraphics[width=.6\textwidth]{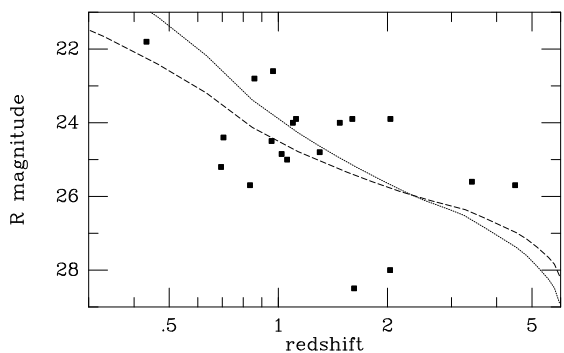}
\end{center}
\caption[]{
The $R$-band Hubble diagram for GRB host galaxies with measured magnitudes
and redshifts, as of June 2001.  The points represent the hosts; typical
error bars are a few tenths of a magnitude.  
The lines represent Bruzual-Charlot galaxy evolution models \cite{bc91}, 
normalised to an $L_*$ galaxy today.  Both models assume Salpeter IMF,
no extinction, galaxy formation redshift $z_{gf} = 6$, and a standard
Friedmann cosmology with $h = 0.65$, $\Omega_0 = 0.2$ and $\Lambda_0 = 0$.
The dotted line corresponds to a model with a constant SFR, and the dashed
line to a model with a SFR declining exponentially with an $e$-folding
time of 9.5 Gyr.  Allowing for extinction would lower the model curves at
higher redshifts; assuming more star formation in the past, and/or a
top-heavy IMF would move them higher.
}
\label{eps1}
\end{figure}

The majority of redshifts so far are from the spectroscopy of host galaxies,
and some are based on the absorption-line systems seen in the spectra of the
afterglows (which are otherwise featureless power-law continua).  Reassuring
overlap exists in some cases; invariably, the highest-$z$ absorption system
corresponds to that of the host galaxy, and has the strongest lines.
In some cases no optical transient (OT) is detected, but a combination of
the X-ray (XT) and radio transient (RT) unambiguously pinpoints the host
galaxy.  
A new method for obtaining redshifts may come from the X-ray spectroscopy of
afterglows, using the Fe K line at $\sim 6.55$ keV \cite{pcf+99,apv+00,pgg+00},
or the Fe absorption edge at $\sim 9.28$ keV \cite{yno+99,wmkr00,amati00}.
Rapid X-ray spectroscopy of GRB afterglows may become a powerful
tool to understand their physics and origins. 

Are the GRB host galaxies special in some way?  If GRBs are somehow related to
the massive star formation (e.g., \cite{tot97,pac98b}, etc.), it may be
worthwhile to examine their absolute luminosities and star formation rates
(SFR), or spectroscopic properties in general.  This is hard to answer
\cite{kth98,hf99,sch00} 
from their visible ($\sim$ restframe UV) luminosities alone: the observed 
light traces an indeterminate mix of recently formed stars and an older
population, cannot be unambiguously interpreted in terms of either the total
baryonic mass, or the instantaneous SFR.  

The magnitude and redshift distributions of GRB host galaxies are typical for
the normal, faint field galaxies, as are their morphologies \cite{hol01} when
observed with the HST: often compact, sometimes suggestive of a merging system
\cite{sgd+01a}, but that is not unusual for galaxies at comparable redshifts. 
Their redshift distribution is about what is expected for an evolving, normal
field galaxy population at these magnitude levels.  There is an excellent
qualitative correspondence between the observations and simple galaxy evolution
models \cite{maomo98}.  This is further illustrated in Fig. 1, which compares
the observed GRB host magnitudes as a function of redshift with stellar
population synthesis models which are used to describe the evolution of normal
field galaxies.  The observed spread of luminosities is reasonable for a
normal, evolving galaxy luminosity function.

\begin{figure}[b]
\begin{center}
\includegraphics[width=.6\textwidth]{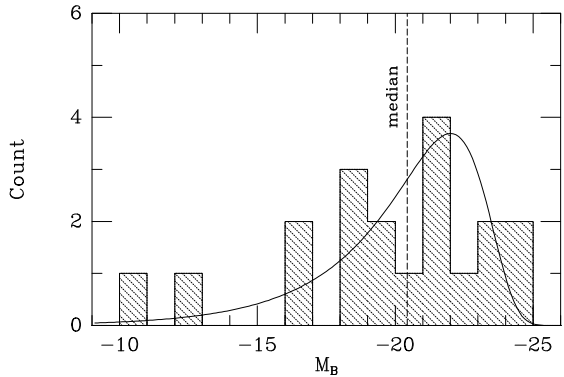}
\end{center}
\caption[]{
Distribution of estimated absolute $B$-band magnitudes for GRB host galaxies
with known redshifts, as of June 2001.  These restframe magnitudes were
computed from the observed $R$-band magnitudes by approximating the galaxy
spectra as $f_\nu \sim \nu^{-1}$ and no additional extinction correction. 
Standard Friedmann cosmology with $H_0 = 65$ km s$^{-1}$ Mpc$^{-1}$, $\Omega_0
= 0.2$, and $\Lambda_0 = 0$ was used.
The dashed line indicates the sample median, $M_B = -20.43$ mag.  
The solid curve is a heuristic model, representing a luminosity-weighted
Schechter function, with $M_* = -23$ mag and $\alpha = -1.6$.  The coice of
parameters is meant to be illustrative, rather than a best fit.
}
\label{eps1}
\end{figure}

Fig. 2 shows the distribution of absolute $B$-band luminosities of GRB hosts
identified to date.  If GRB's follow the luminous mass, then the expected
distribution would be approximated by the luminosity-weighted galaxy luminosity
function for the appropriate redshifts.
The hosts span a wide range of luminosities, with the sample median very close
to a present-day average ($L_*$) galaxy. 
The observed distribution can indeed be reasonably described by an evolved
Schechter luminosity function, with an overall brightening and a steepening on
the low-luminosity end. 
The interpretation of this result is complex: the observed light reflects an
unknown combination of the unobscured fraction of recent star formation
(especially in the high-$z$ galaxies, where we observe the restframe UV
continuum) and the stellar populations created up to that point. 

This is in a qualitative agreement with studies of field galaxy evolution 
\cite{lth+95,ellis97}, which indicate both a brightening of $M_*$, and a
steepening of $\alpha$, relative to $z \sim 0$ galaxy samples. 
The apparent frequent occurence of ``subluminous'' hosts is not surprising,
since most star forming activity at $z \sim 1$ is in starbursting dwarfs
(corresponding to the steepening of the power-law tail of the GLF).
Thus the GRB hosts seem to be representative of the normal, star-forming
field galaxy population at comparable redshifts, and so far there is no
evidence for any significant systematic differences between them.

One could also speculate that lower luminosity galaxies may on average have
lower metallicities, where the mean extinction may be lower, making them 
easier to detect, or whose stellar IMF may be biased towards more massive stars
(this is $highly$ speculative).

Within the host galaxies, the distribution of GRB-host offsets follows closely 
the light distribution \cite{bdk01,bk01}, 
which is roughly proportional to the density of star formation (especially for
the high-$z$ galaxies).  It is thus fully consistent with a progenitor
population associated with the sites of massive star formation.

\begin{figure}[b]
\begin{center}
\includegraphics[width=.5\textwidth]{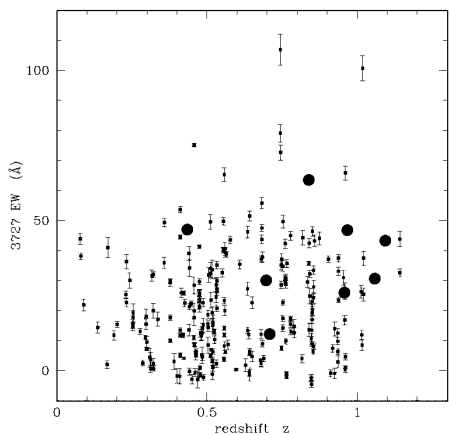}
\end{center}
\caption[]{
Restframe equivalent widths of [O II] 3727 line emission for a
magnitude-limited sample of field galaxies from Hogg et al. \cite{hcbp98}
(small squares), and a sample of GRB hosts (large circles).  There is a hint
that the GRB hosts tend to have higher equivalent widths at comparable
redshifts, relative to this field galaxy sample.  However, the GRB hosts tend
to be fainter on average, as this field sample is magnitude-limited at $R = 23$
mag.  The apparent difference in mean equivalent widths may be a combination of
sample selection and evolution effects, rather than an evidence for enhanced
SFR per unit mass for GRB hosts. 
}
\label{eps2}
\end{figure}

Spectroscopic measurements provide direct estimates of the recent, massive
SFR in GRB hosts.  Most of them are based on 
the luminosity of the [O II] 3727 doublet \cite{ken98}, 
the luminosity of the UV continuum at $\lambda_{rest} = 2800$ \AA\ \cite{mpd98},
in one case so far \cite{kdr+98}
from the luminosity of Ly$\alpha$ 1216 line \cite{kdr+98},
and in one case \cite{dkb+98}
from the luminosity of Balmer lines \cite{ken98}.
All of these estimators are susceptible to the internal extinction and its
geometry, and have an intrinsic scatter of at least 30\%.
The observed SFR's range from a few tenths to a few $M_\odot$ yr$^{-1}$,
again typical for the normal field galaxy population at comparable redshifts.

Equivalent widths of the [O II] 3727 doublet in GRB hosts, which may provide
a crude measure of the SFR per unit luminosity (and a worse measure of the
SFR per unit mass), are on average somewhat higher \cite{sgd+01c}
than those observed in magnitude-limited field galaxy samples at comparable
redshifts \cite{hcbp98}.
This is illustrated in Fig. 3.
A larger sample of GRB hosts, and a good comparison sample, matched both 
in redshift and magnitude range, are necessary before any solid conclusions can
be drawn from this apparent difference. 

One intriguing hint comes from the flux ratios of [Ne III] 3869 to
[O II] 3727 lines: they are on average a factor of 4 to 5 higher in GRB
hosts than in star forming galaxies at low redshifts.  These strong [Ne III] 
require photoionization by massive stars in hot H II regions, and may represent
an indirect evidence linking GRBs with massive star formation.

The interpretation of the luminosities and observed star formation rates is
vastly complicated by the unknown amount and geometry of extinction.  The
observed quantities (in the visible) trace only the unobscured stellar
component, or the components seen through optically thin dust.  Any 
stellar and star formation components hidden by optically thick dust cannot
be estimated at all from these data, and require radio and sub-mm observations.
Thus, for example, optical colors of GRB hosts cannot be used to make any
meaningful statements about their net star formation activity.  The broad-band
optical colors of GRB hosts are not distinguishable from those of normal
field galaxies at comparable magnitudes and redshifts \cite{bdk01,sfct+01}.

Already within months of the first detections of GRB afterglows, no OT's were
found associated with some well-localised bursts despite deep and rapid
searches; the prototype ``dark burst'' was GRB 970828 \cite{sgd+01b}.
Perhaps the most likely explanation for the non-detections of OT's when
sufficiently deep and prompt searches are made is that they are obscured by
dust in their host galaxies.  This is an obvious culprit if indeed GRBs are
associated with massive star formation.

Support for this idea also comes from detections of RTs without OTs, including
GRB 970828, 990506, and possibly also 981226 (see \cite{tbf+00,fbg+00}).
Dust reddening has been detected directly in some OTs 
(e.g., \cite{rkf+98,bfk+98,dkb+98} etc.);
however, this only covers OTs seen through optically thin dust, and there
must be others, hidden by optically thick dust.
An especially dramatic case was the RT \cite{tfk+98} and 
IR transient \cite{lg+98}
associated with GRB 980329.  We thus know that at least some GRB OTs must be
obscured by dust. 

The census of OT detections for well-localised bursts can thus provide a
completely new and independent estimate of the mean obscured star formation
fraction in the universe.  Recall that GRBs are now detected out to 
$z \sim 4.5$ and that there is no correlation of the observed fluence with 
the redshift \cite{mg9},
so that they are, at least to a first approximation, good probes of the star
formation over the observable universe.  

As of mid-June 2001, there have been $\sim 52 \pm 5$ adequately deep and rapid
searches for OTs from well-localised GRBs. 
We define ``adequate searches'' as reaching at least to $R \sim 20$ mag within
less than a day from the burst, and/or to at least to $R \sim 23 - 24$ mag
within 2 or 3 days; this is a purely heuristic, operational definition.  The
uncertainty comes from the subjective judgement of whether the searches really
did go as deep and as fast, and whether the field was at a sufficiently low
Galactic latitude to cause concerns about the foreground extinction and
confusion by Galactic stars. 
Out of those, $\sim 27 \pm 2$ OTs were found (the uncertainty being due to the
questionable nature of some candidates).
Some OTs may have been missed due to an intrinsically low flux, an unusually
rapid decline rate, or very high redshifts (so that the brightness in the
commonly used $BVR$ bands would be affected by the intergalactic absorption).
Thus the $maximum$ fraction of all OTs (and therefore massive star formation)
hidden by the dust is $(48 \pm 8)$\%.

This is a remarkable result.  It broadly agrees with the estimates that there
is roughly an equal amount of energy in the diffuse optical and FIR backgrounds
(see, e.g., \cite{mad99}).  This is contrary to some claims in the literature
which suggest that the fraction of the obscured star formation was much higher
at high redshifts.  Recall also that the fractions of the obscured and
unobscured star formation in the local universe are comparable.  GRBs can
therefore provide a valuable new constraint on the history of star formation
in the universe. 

There is one possible loophole in this argument: GRBs may be able to destroy
the dust in their immediate vicinity (up to $\sim 10$ pc?) \cite{wd00,gw00},
and if the rest of the optical path through their hosts ($\sim$ kpc scale?)
was dust-free, OTs would become visible.  Such a geometrical arrangement may
be unlikely in most cases, and our argument probably still applies.
Further support for the use of GRBs as probes of obscured star formation in
distant galaxies comes from radio and sub-mm detections of the hosts
\cite{fr+01b,bkf01}.

\noindent{\bf Acknowledgments:}~
We wish to acknowledge the efforts of numerous collaborators worldwide, and the
expert help of the staff of Palomar, Keck, and VLA observatories and STScI. 
This work was supported in part by grants from the NSF, NASA,
and private foundations to SRK, SGD, FAH, and RS, Fairchild Fellowships to RS
and TJG, Hubble Fellowship to DER, and Hertz Fellowship to JSB.

\end{document}